\documentclass[preprint]{aastex}

\shorttitle{Halo Population of NGC~5128}
\shortauthors{Marleau et al.}

\newcommand{\mydeg}{\mbox{$^\circ$}}

\newcommand{\myhrs}{\mbox{$^{\rm h}$}}
\newcommand{\mymins}{\mbox{$^{\rm m}$}}

\newcommand{\mysecsd}{\mbox{$^{\rm s} \!\! .$}}
\newcommand{\littleprime}{\ifmmode{\scriptscriptstyle \prime }
\else{\hbox{$\scriptscriptstyle \prime$ }}\fi}
\newcommand{\myarcsec}{\raise .9ex \hbox{\littleprime\hskip-3pt\littleprime}}
\newcommand{\myarcmin}{\raise .9ex \hbox{\littleprime}}
\newcommand{\myarcsecpoint}{\hbox to 1pt{}\rlap{\myarcsec}.\hbox to 2pt{}}
\newcommand{\myarcminpoint}{\hbox to 1pt{}\rlap{\myarcmin}.\hbox to 2pt{}}
\newcommand{\simlt}{\mathrel{\spose{\lower 3pt\hbox{$\mathchar"218$}}
     \raise 2.0pt\hbox{$\mathchar"13C$}}}
\newcommand{\simgt}{\mathrel{\spose{\lower 3pt\hbox{$\mathchar"218$}}
     \raise 2.0pt\hbox{$\mathchar"13E$}}}

\begin{document}

\title{The Nature of the Halo Population of NGC~5128 \\
Resolved with NICMOS on the {\it Hubble Space Telescope}}

\author{Francine R. Marleau}
\affil{Institute of Astronomy, University of Cambridge, Madingley Road, \\
Cambridge CB3 0HA, England, U.K.}
\author{James R. Graham}
\affil{Department of Astronomy, University of California, Berkeley, \\
Campbell Hall, Berkeley, CA 94720, USA}
\author{Michael C. Liu}
\affil{Department of Astronomy, University of California, Berkeley, \\
Campbell Hall, Berkeley, CA 94720, USA}
\author{St\'ephane Charlot}
\affil{Institut d'Astrophysique du CNRS, 98 bis Boulevard Arago, \\
75014 Paris, France}

\begin{abstract}
We present the first infrared (IR) color-magnitude diagram (CMD) for the
halo of a giant elliptical galaxy.  The CMD for the stars in the halo of
NGC~5128 (Centaurus~A) was constructed from HST NICMOS observations of
the WFPC2 CHIP-3 field of Soria et al. (1996) to a 50\% completeness
magnitude limit of [F160W]=23.8. This field is located at a distance of
08\myarcmin50\myarcsec ($\sim9$~kpc) south of the center of the galaxy.
The luminosity function (LF) shows a marked discontinuity at
[F160W]$\approx$20.0. This is $1-2$ mag above the tip of the red giant
branch (TRGB) expected for an old population ($\sim12$~Gyr) at the
distance modulus of NGC~5128. We propose that the majority of stars
above the TRGB have intermediate ages ($\sim2$~Gyr), in agreement with
the WFPC2 observations of Soria et al. (1996). Five stars with
magnitudes brighter than the LF discontinuity are most probably due to
Galactic contamination. The weighted average of the mean giant branch
color above our 50\% completeness limit is
[F110W]$-$[F160W]=1.22$\pm0.08$ with a dispersion of 0.19 mag. From our
artificial-star experiments we determine that the observed spread in
color is real, suggesting a real spread in metallicity. We estimate the
lower and upper bounds of the stellar metallicity range by comparisons
with observations of Galactic star clusters and theoretical
isochrones. Assuming an old population, we find that, in the halo field
of NGC~5128 we surveyed, stars have metallicities ranging from roughly
1\% of solar at the blue end of the color spread to roughly solar at the
red end, with a mean of [Fe/H]=$-0.76$ and a dispersion of 0.44 dex.
\end{abstract}


\section{Introduction}

Individual stars are the visible building blocks of galaxies and direct
tracers of the galaxy formation process. Massive elliptical galaxies are
believed to contain the majority of the oldest stars in the Universe
\citep{tinsley76}. Two main scenarios have been proposed for the
formation of these galaxies. In the traditional scenario of single
``monolithic'' collapse, the most massive early-type (E and S0) galaxies
form at early times ($z\gtrsim2$) and evolve with little or no star
formation thereafter \citep{tinsley80, bruzual80,koo81,shanks84, king85,
yoshii88, guiderdoni90}. In hierarchical models of galaxy formation,
massive galaxies assemble later ($z\lesssim2$) from mergers of smaller
subunits \citep{white78,white91,lacey93,kauffmann93, cole94,
somerville97}. At least some elliptical galaxies show signatures of
intermediate-age stars in addition to an old stellar population
\citep{worthey92}. Dynamical studies also suggest that the pressure
support of stellar populations in elliptical galaxies could result from
mergers \citep{toomre72,hibbard94}. Moreover, detailed observations of
some elliptical galaxies reveal morphological and kinematic signs of a
past merging event. These range from the observations of
proto-elliptical merger remnants like NGC~7252 to evidence of
counter-rotating or otherwise decoupled cores for nearby ellipticals
(star-star or star-gas, Bertola 1997). The dominance of old stars and
evidence of merging in elliptical galaxies can be understood
simultaneously if the youngest stars contribute only a small fraction of
the observed integrated light.  In fact, \citet{silva98} show that
nearby elliptical galaxies with morphological signatures of recent
merger activity have near-IR colors similar to those of galaxies not
showing signs of mergers. They conclude that intermediate-age ($1-3$
Gyr) stars contribute at most 10\%$-$15\% of the total stellar mass in
galaxies with recent merger activity in their sample.

Resolving individual stars in elliptical galaxies has become feasible
only recently with the advent of the Hubble Space Telescope (HST), since
no suitable examples are near enough to be observed from the ground
(this will change with the application of adaptive optics systems to
large ground-based telescopes).  Such detailed information about the
stellar content of elliptical galaxies can help us reconstruct their
star formation history, and hence, constrain their process of formation.
With HST, it is possible to resolve the population of the massive
elliptical galaxy NGC~5128 (Centaurus~A) due to its proximity to us.
NGC~5128 is part of a group of 25 galaxies composed mainly of dwarf
galaxies extending over about 25 degree on the sky (see review by Israel
1998).  Its distance estimate relies upon different measurement methods.
A distance modulus of $(m-M)_0=27.53 \pm 0.25$ \citep{tonry90} is
derived from the globular cluster luminosity function (LF) of
\citet{harris86}. The planetary nebula luminosity function yields
$(m-M)_0=27.73\pm0.14$ \citep{hui93}. The distance modulus of
$(m-M)_0=27.48\pm0.06$ measured by \citet{tonry90} from surface
brightness fluctuations is revised to $(m-M)_0=27.71\pm0.10$ by
\citet{israel98} using the results of \citet{tonry91} (the more recent
results of \citet{tonry97} yield a higher revised value of
$(m-M)_0=28.18\pm0.07$).  An estimate of $(m-M)_0=27.72\pm0.20$ is
derived from the magnitude of the tip of the red giant branch (TRGB) for
stars in the halo observed with HST WFPC2 by Soria et al. (1996;
hereafter SMW96).  A more recent measurement by Harris et al. (1999;
hereafter HHP99) comes from the magnitude of the TRGB for stars in
another halo field observed with HST WFPC2 (these data reach $\sim3$
magnitudes down the RGB).  They find a distance modulus of
$(m-M)_0=27.98\pm0.15$, or $D=3.9\pm0.3$ Mpc.  The weighted average of
distance moduli of $(m-M)_0=27.75\pm0.06$ is adopted throughout this
paper and corresponds to a distance of $D=3.5\pm0.1$ Mpc.  At this
distance, 1 arcmin corresponds to 1018~pc.

NGC~5128 is a clear case of an elliptical galaxy showing signs of past
merger activity.  It is one of the largest known radio galaxies (500
$\times$ 250 kpc wide) and a massive disk of gas, young stars, and dust
is embedded in its center.  Within a radius of about 18\myarcmin from
the nucleus, the galaxy shows optical shell structures made up of old
disk stars and associated H\,{\sc i} shells \citep{schiminovich94}.
This suggests that NGC~5128 might have experienced more than just one
merger in its past \citep{weil96}.  The H\,{\sc i} shells detected in
the outer part of the galaxy seem to show signs of recent star formation
\citep{graham98}.  The halo globular clusters in NGC~5128 (region in
radius $R<$~24\arcmin) have a mean metallicity of [Fe/H]=$-0.8\pm0.2$
(0.5 dex higher than their Milky Way counterparts; Harris et al. 1992)
and show a bimodal distribution in color, with peaks at
[Fe/H]$\simeq-1.1$ and $-0.3$ ($R>$ 4\arcmin; Harris et al. 1992), an
effect commonly associated with a merging event.  \citet{jablonka96}
find no object with a metallicity higher than solar in a similar
sample. The globular clusters in the inner 3~kpc are more metal rich
with $-0.6\leq$[Fe/H]$\leq+0.1$ than in the outer regions and show signs
of a metallicity gradient \citep{jablonka96, minniti96, alonso97}.
Recent HST WFPC2 observations yield a mean value of [Fe/H]$>-0.9$
(SMW96) and [Fe/H]=$-0.41$ (HHP99) for red giant branch stars in the
halo of NGC~5128.

The age of the current burst of star formation in the disk of NGC~5128
is typically a few times $10^7$ years \citep{vandenbergh76,dufour79}.
In the halo, the presence of $\sim200$ stars found to be brighter than
the TRGB prompted SMW96 to suggest the presence of an intermediate-age
population of $\sim5$~Gyr, making up at most 10\% in number of the total
halo population.  Dynamical estimates based on the model of a merger of
a late-type galaxy of mass a few times $10^{10} M_{\odot}$ with NGC~5128
suggest a more recent merging timescale of a few hundred million years
\citep{tubbs80,malin83}.  The total mass of the galaxy estimated from
velocity dispersion measurements of the planetary nebula system is
$M=4\pm1 \times 10^{11} M_{\odot}$, with half of it estimated to be due
to dark matter \citep{mathieu96}.

We present in this paper the first IR color-magnitude diagram (CMD) for
the halo of a giant elliptical galaxy.  Our NICMOS observations of the
halo of NGC~5128 probe a range of $\sim4$ magnitudes down the luminosity
function to our 50\% completeness limit.  Section~2 presents the details
of the NICMOS observations we obtained in August 1998.  The data
analysis using the stellar photometry package DAOPHOT is described in
Section~3.  The importance of doing artificial-stars experiments is
emphasized in Section~4, where the completeness functions are presented.
The luminosity functions and discontinuities are derived in Section~5.
Section~6 presents the first IR CMD of the halo of NGC~5128.  This
section is divided into four sub-sections discussing the total
uncertainties, deriving an estimate of the metallicity spread of the
halo stars, and discussing the nature of the bright stars above the TRGB
of an old population.  A summary of our results and conclusions appears
in Section~7.  The more technical details of the magnitude system
transformations are given in Appendix~A.

\section{Observations}

Images of a field in the halo of NGC~5128 were taken on 1998 August 31
with the NIC1 and NIC2 camera of the Near-Infrared and Multiobject
Spectrometer (NICMOS) \citep{thompson98} on board HST.  The spacecraft
pointing was chosen to image the existing WFPC2 CHIP-3 field of SMW96,
at a distance of 08\myarcmin50\myarcsec from the nucleus, with the
NICMOS cameras (see Figure~\ref{ngc5128_dss_15}).  We chose our
positions of NIC1 and NIC2 so they would lie inside the WFPC2 CHIP-3
field of view of SMW96 for any unrestricted value of position angle.
Based on our adopted distance of 3.5~Mpc, this corresponds to a distance
of $\sim9$~kpc south of the center of the galaxy.  The geometrical
center of NIC2 was at position $\alpha=+13$\myhrs25\mymins24\mysecsd323,
$\delta=-43$\mydeg09\myarcmin58\myarcsecpoint53 (J2000) and the NIC1
observations were obtained in parallel.  The NIC2 detector has a field
of view of 19\myarcsecpoint2 $\times$ 19\myarcsecpoint2 with 256 pixels
on a side and 0\myarcsecpoint075 per pixel.  A total of 8192~s (3
orbits) of integration time were obtained, 5376~s in the F160W filter
and 2816~s in the F110W filter.  The observations were taken in a four
dither positions mode with offsets of (0,0), (0,15.5), (15.5,15.5), and
(15.5,0) pixels, corresponding to a maximum shift of 1\myarcsecpoint17.
This enables the replacement of a bad pixel in one frame with the
average of good pixels from dithered frames.  In addition, short
exposures were taken at the beginning of each orbit in the MULTIACCUM
mode to reduce the cosmic rays persistence effects \citep{najita98}.
All the observations were taken in the MULTIACCUM mode with the SPARS64
sequence.

A basic image reduction was done by STScI using the standard NICMOS
pipeline procedure called CALNICA which performs bias subtraction,
dark-count correction, and flat-fielding \citep{mackenty97}.  Our own
subsequent data reduction with IRAF\footnote{IRAF is distributed by the
National Optical Astronomy Observatories, which are operated by the
Association of Universities for Research in Astronomy, Inc., under
cooperative agreement with the National Science Foundation.} consisted
of masking bad pixels, correcting for a constant level offset between
the four quadrants of the NIC2 camera, correcting for background
variations by subtracting a median sky image and adding back a constant
sky level, and averaging the images.  Each resulting NIC2 mosaic covers
a field of view of 20\myarcsecpoint4 $\times$ 20\myarcsecpoint4.  The
final combined NIC2-F110W and -F160W images of our field are shown in
Figure~\ref{ngc5128_nic2}.

\section{Stellar Photometry}

Photometry was obtained for our NICMOS images by using the automated
star-detection algorithm DAOPHOT \citep{stetson87,stetson92}.  The data
were initially processed using the subroutine DAOFIND with a
conservative detection threshold of $5\sigma$ above the local background
level.  The PSFs were derived by using many well-exposed, isolated stars
sampling the frame uniformly.  The full width at half-maximum (FWHM) of
the F160W and F110W PSF profiles were measured to be equal to 1.7 and
1.2 pixel, respectively.  The stellar photometry was accomplished by
processing our images using the subroutine ALLSTAR once, and then
another time on the first residual image.  The fitting was done on
pixels within a fitting radius of the centroid of a star equal to the
FWHM of the PSF.  The 3-parameter least-square fit to the star
determines the position of the center (x,y), the brightness and its
standard deviation.  We measured magnitudes for 971 stars in the F110W
image and 1321 stars in the F160W image.

The residual images generated by DAOPHOT show that stars near the edges
of the mosaics suffer from bad PSF fits (their distance from the edge is
of the order of their FWHM).  Our final sample contains stars selected
so that they are not too close to the edge of the mosaic, i.e., no less
than 3 pixels away from the edge.  We also applied a stellar fit
$\chi^2$ cut to our sample of stars.  The cut was set to a $\chi^2$ of
2.6, given that the probability of exceeding that value of $\chi^2$ is
5\%.  The $\chi^2$ test was done only on the F160W image for which the
PSF is well sampled.  In the case of the under-sampled F110W image for
which the PSF is almost a Dirac delta function, $\chi^2$ provides a poor
discriminant because it is possible to fit every pixel that contains a
signal.  The [F160W] $\chi^2$ cut eliminated 215 stars of which 3 were
on the right hand side of the red envelope of the normal giants in our
CMD, leaving a single extremely red star at
[F110W]$-$[F160W]=2.5$\pm0.1$ (see Section~6 and Figure~\ref{cm_mean}).

The instrumental magnitudes were transformed into the Vega-based system.
Since the photometric keywords needed to transform countrates into
magnitudes refer to a nominal infinite aperture, given in the HST Data
Handbook as 1.15 times the flux in a 0.5 arcsec radius aperture, the
measured countrates have to be corrected accordingly.  This is done by
first correcting the measured PSF-fitting photometry to the
0\myarcsecpoint13 radius aperture photometry (a 0\myarcsecpoint13 or 1.7
pixel radius aperture, of the order of the FWHM of the F160W PSF, was
used to calibrate the PSF flux in both filters). Secondly, in order to
estimate the aperture correction, we used both our observed PSF and the
artificial PSF generated by the Space Telescope package TINYTIM
\citep{krist93}.  We found for our choice of aperture radius of
0\myarcsecpoint13 that the fraction of PSF-fitting photometry to
0\myarcsecpoint5 radius aperture was 60\% for the F110W filter and 56\%
for the F160W filter.  Putting in the final 1.15 correction factor, we
calculated a correction to brighter magnitudes of 0.71 and 0.77,
respectively.  The photometric calibration of
PHOTFNU~=~$2.190\times10^{-6}$~Jy~sec/DN and ZP(Vega)~=~1083~Jy (HST
Data Handbook; updated values from Rieke~1999) produced the NIC2-F160W
zeropoint in the Vega-based system of 21.74.  After including the
aperture correction, the Vega-based magnitudes for stars in our F160W
image were therefore computed by using mag
(Vega-based)~=~$20.97-2.5$log(counts(e-/s)) (gain is 5.0 e-/DN).  The
F110W zeropoint was calculated to be 21.64, given
PHOTFNU~=~$2.031\times10^{-6}$~Jy~sec/DN, ZP(Vega)~=~1775~Jy and the
aperture correction given above.  The systematic errors in magnitude and
color measurements are estimated to be less than 0.05~mag (Calzetti et
al.~1999; see Colina and Rieke 1997 for more detail).

Colors were calculated for the stars in the NICMOS images in the
following way.  Each color image was analyzed separately.  The
identification of a star in both images was done by requesting that (1)
the two positions agreed to a specific tolerance (matching) radius, and
(2) the star assigned was the closest uniquely assigned star in either
color image.  A total of 666 stars (out of the sample of 1087 stars in
F160W and 941 stars in F110W) were identified in both filters using a
maximum matching radius $r_m=1$ pixel.  This maximum matching radius
was chosen because the number of stars matched show a clear cut at that
radius, as seen in Figure~\ref{matches} where we allowed the matching
radius to be as large as $r_m=5$ pixels.

To justify our choice of the maximum matching radius and illustrate that
the peak of the distribution lies between $0.1-0.2$ pixel, we modeled
the number of stars with a match between the F110W and F160W images.
The model is made up of a Gaussian component that reflects the
uncertainty in the centroid of each image, due to undersampling in the
F110W image, photon statistics, and the effects of crowding.  To this we
add a constant background probability of making a false match with an
un-associated star.  Hence the number of stars matching, $N$, between
radii $r$ and $r+dr$ should follow:

\[ N(r) \; dr = N_{true} \; \left( \frac{r}{\sigma^2} \right) \; 
e^{-0.5 ( \frac{r}{\sigma})^2} \; dr \; + \; 2 \; \pi \; r \; N_\ast \; dr, \]

\noindent
where $N_{true}$ is the number of true matches, $N_\ast$ is the number
of un-associated stars per unit area, and $\sigma$ represents the
two-dimensional relative error in star positions between the two frames.
The best fit curve is shown over-plotted on the histogram of matching
radius in Figure~\ref{matches}. This implies that the net
two-dimensional root-mean-square (rms) registration error between the
F110W and F160W frames is 0.19 pixel, and that a natural choice of
cut-off is at a radius of 1 pixel which is equivalent to $5\sigma$. The
excess matches above the fit between radii of $0.6-0.9$ pixels is due to
genuine matches of faint stars with poorly determined centroids.  The
false match rate implies that within our chosen cutoff of 1 pixel, 3\%
of our stars are matched with the wrong counterpart.

\section{Artificial-Star Tests}

In the absence of other systematic errors, the photometric errors depend
only on photon statistics (from source and background) and detector
noise.  Simulations we ran prior to our observations showed that in
order to be limited by these errors and not be affected by crowding, we
needed to observe a region with a surface brightness no higher than 21.0
mag/arcsec$^2$ in the F160W filter (assuming a Baade's window LF).  This
is consistent with analytic estimates of the effects of crowding
\citep{renzini98}.  In the simulations, we were able to recover the
input luminosity function down to [F160W]$\simeq23.5$ with photometric
accuracy to the 10\% level.  To assess the errors associated with doing
photometry in our field, we simulated artificial stars in our NICMOS
images.

\begin{center}
\begin{tabular}{c}
TABLE 1 \\
\small{COMPLETENESS}
\end{tabular}
\end{center}
\vspace{-0.75cm}
\begin{center}
\begin{tabular}{ccc} 
\hline \hline
 Magnitude & Completeness (\%) & Completeness (\%) \\
           & [F110W]      & [F160W] \\
\hline 
 20.0 & 100 & 100 \\
 20.5 & 100 &  99 \\
 21.0 & 100 &  98 \\
 21.5 &  98 &  98 \\
 22.0 &  99 &  98 \\
 22.5 &  98 &  98 \\
 23.0 &  98 &  94 \\
 23.5 &  95 &  78 \\
 24.0 &  87 &  33 \\
 24.5 &  52 &   5 \\
 25.0 &  17 &   3 \\
 25.5 &   4 &   2 \\
 26.0 &   2 &   2 \\
 26.5 &   2 &   2 \\
\hline
\end{tabular}
\end{center}

Completeness tests were performed by adding artificial stars to each
individual color image.  We simulated stars with F110W and F160W
magnitudes between 20.0 and 26.5, at a 0.1 mag interval.  Each
simulation consisted of adding 132 stars to the real image,
corresponding to 10\% in number of the stars recovered from the F160W
image.  The number of artificial stars was chosen to be large enough to
compile accurate statistics on incompleteness and photometric errors,
and small enough to increase the crowding negligibly.  The positions of
the added stars on the images were randomly chosen but identical for
both the F110W and F160W images so that colors could be measured.  The
frames were then processed in a manner identical to the original data.
The F110W and F160W completeness functions measured from these
artificial-star tests are shown in Figure~\ref{completeness} and listed
in Table~1.  The 50\% completeness level occurs at [F110W]=24.5 and
[F160W]=23.8, respectively.

\section{The Luminosity Functions}

Our NICMOS observations probe $\sim4$ magnitudes below the tip of the
luminosity function down to our 50\% completeness limit.  The NICMOS
data are less crowded by bright stars than the WFPC2 data since the IR
luminosity functions tend to be steeper and so there are fewer bright
stars.  Figure~\ref{lf_all_match} shows the luminosity functions for
stars detected in the F110W and F160W image respectively, taking into
account the edge and $\chi^2$ cuts.  Also shown in
Figure~\ref{lf_all_match} and listed in Table~2 are the luminosity
functions for each respective color image obtained after applying the
matching criterion ($r_m=1$ pixel).  The red F160W stars that were not
matched/detected in the F110W image contribute to the F160W LF only at
faint magnitudes ([F160W]$\gtrsim22.0$).  

The matched F160W LF displayed in Figure~\ref{lf_all_match} clearly
shows a discontinuity in the number of stars at [F160W]$\approx$20.0.
Although no other clear discontinuity is visible at fainter magnitudes,
the slope of the counts changes slightly at [F160W]$\approx$21.0.  As
will be discussed in detail in Section 6, these discontinuities are
thought to be associated with the TRGB of an intermediate-age and old
population in the halo of NGC~5128.  In the IR, the presence of an
intermediate-age population and an old population will be blended and
make the TRGB of the old population difficult to detect while the
younger population will be clearly brighter and more visible.  Note that
the stars brighter than the [F160W]$\approx$20.0 discontinuity are most
probably due to Galactic contamination (see Section~6.3 for discussion).

Given that no IR surface brightness measurements for the halo of
NGC~5128 can be found in the literature, it is only possible to estimate
this based on the V-band measurements of \citet{vandenbergh76}.  The
V-band surface brightness at the radial distance of our field is
measured to be 23.2 mag/arcsec$^2$.  This corresponds to a [F160W]
surface brightness value of 20.2$\pm0.1$ mag/arcsec$^2$, assuming the
color transformation $V-H=3.1\pm0.1$ \citep{persson79} and using the
magnitude system transformations given in Appendix~A.  This mean color
and standard deviation are calculated based on the distribution in $V-H$
of the \citet{persson79} field ellipticals.  The total light in our
resolved population that passes the edge and $\chi^2$ cuts averages to
21.3 mag/arcsec$^2$.  Hence, $\sim40$\% of the light is resolved into
the stars appearing in the [F160W] LF (see Figure~\ref{lf_all_match};
top panel).  This result is consistent with a Baade's window luminosity
function (used in our simulations) and confirms that we are not confused
by crowding for [F160W]$\lesssim23.5$.

\section{IR Color Dispersion}

The NICMOS CMD for NGC~5128 is shown in Figure~\ref{cm_mean}.  An
electronic version of the photometry table may be obtained on request
from the first author.  The foreground reddening in the direction of our
NICMOS field is $E(B-V)=0.11\pm0.02$ \citep{frogel84,harris92}. For
$R_V=3.1$, this corresponds to $A_J=0.10$, $A_H=0.06$ and $E(J-H)=0.04$
\citep{rieke85}.  Also shown in this figure is the mean color and
standard deviation measured at each magnitude bin.  The color histograms
and Gaussian fits from which these means and standard deviations were
computed are shown in Figure~\ref{color_gauss}.  The reader is referred
to Table~2 for the list of statistical measurements computed for the
real data as well as for the artificial-stars tests.  The weighted
average of the mean giant branch color above our 50\% completeness limit
is [F110W]$-$[F160W]=1.22$\pm0.08$ ($(J-H)_{CIT}=0.78$) with a
dispersion of 0.19 mag.

\subsection{Discussion of Uncertainties}

In order to estimate the metallicity spread in our IR-selected sample of
halo stars, we first need to check whether the spread in color we detect
in our data is real.  We can then use the Bruzual and Charlot (2000;
hereafter BC00) isochrones, described in detail in Section~6.2, to
estimate the metallicity spread associated with the observed real color
spread.

We begin by comparing the observed spread with the total uncertainties
(statistical+systematic errors) calculated from the artificial-star
tests.  The error in color cannot simply be measured by adding in
quadrature the uncertainty in [F110W] and [F160W] because the covariance
of these errors is not zero; i.e., the errors are correlated. This
occurs because the photometric errors are not random but associated with
systematic errors that occur in the photometry of crowded fields; faint
stars next to brighter ones will have correlated errors in their F160W
and F110W photometry.  The photometric errors for the [F160W] magnitudes
do not appear in Figure~\ref{cm_mean} as they are smaller than the
symbols used to plot the points.  The mean of the DAOPHOT [F160W]
photometric errors for our artificial-star tests are added up in
quadrature and listed in Table~2.  The artificial-star tests total
uncertainty measurements become larger than the DAOPHOT photometric
errors only in the faintest magnitude bin where the systematic errors
(mostly due to crowding of faint stars) become dominant.

The standard deviations in mean color for the seven magnitude bins, as
listed in Table~2 and shown in Figure~\ref{cm_mean}, can be compared
with the total uncertainties measured from the artificial-stars tests.
The observed spread is larger than the total uncertainty in all bins
except the faintest one near our 50\% completeness limit.  For the
faintest magnitude bin, we are only $0.3-0.8$ magnitude above our 50\%
completeness limit.  For the three brightest bins, the observed spread
is larger than the total uncertainty by at least a factor of 1.5--3.5.
Assuming that the total uncertainty and intrinsic color spread add in
quadrature, the observed rms for the magnitude bin [F160W]=22.0--22.5 of
0.20$\pm0.02$~mag implies a {\it real} color spread of
0.11$\pm0.02$~mag.  We find for the 12~Gyr isochrones of BC00 (see
Section~6.2) that at $M_{[F160W]}=-5.5$:

\[ \frac{d[Fe/H]}{d([F160W]-[F110W])} = 2.03 \; \rm{dex/mag}. \]

\noindent
This means that a 0.1 mag error in color corresponds to a 0.2 dex error
in metallicity.  Hence, the real color spread of 0.11$\pm0.02$~mag
corresponds to a rms for [Fe/H] of 0.22$\pm0.04$~dex.  The FWHM of the
metallicity distribution is then 0.5$\pm0.1$~dex.

\begin{center}
\begin{tabular}{c}
TABLE 2 \\
\small{LUMINOSITY FUNCTIONS AND COMPARISON BETWEEN} \\
\small{OBSERVED [F110W]$-$[F160W] COLOR SPREAD AND TOTAL UNCERTAINTY} 
\end{tabular}
\end{center}
\vspace{-0.6cm}
\begin{center}
\begin{tabular}{cccccccc} 
\hline \hline 
 [F160W] & N$_{F160W}$ & Observed   & Observed  & Artificial & Artificial & DAOPHOT     & N$_{F110W}$ \\
 mag bin & ($\chi^2 \leq 2.6)$ & mean color & rms color & mean color & rms color
 & rms color error & \\ 
\hline 
 17.0$-$17.5 &   1 & \nodata & \nodata & \nodata & \nodata & &   0 \\
 17.5$-$18.0 &   0 & \nodata & \nodata & \nodata & \nodata & &   1 \\
 18.0$-$18.5 &   0 & \nodata & \nodata & \nodata & \nodata & &   0 \\
 18.5$-$19.0 &   1 & \nodata & \nodata & \nodata & \nodata & &   0 \\
 19.0$-$19.5 &   2 & \nodata & \nodata & \nodata & \nodata & &   0 \\
 19.5$-$20.0 &   1 & \nodata & \nodata & \nodata & \nodata & &   1 \\
 20.0$-$20.5 &   6 & \nodata & \nodata & \nodata & \nodata & &   2 \\
 20.5$-$21.0 &  24 & 1.25 & 0.17$\pm0.02$ & 1.3 & 0.05 & 0.10 &   1 \\
 21.0$-$21.5 &  39 & 1.24 & 0.14$\pm0.02$ & 1.3 & 0.06 & 0.11 &   1 \\
 21.5$-$22.0 &  95 & 1.27 & 0.18$\pm0.02$ & 1.3 & 0.11 & 0.11 &  20 \\
 22.0$-$22.5 & 163 & 1.24 & 0.20$\pm0.02$ & 1.2 & 0.17 & 0.13 &  37 \\
 22.5$-$23.0 & 165 & 1.18 & 0.23$\pm0.01$ & 1.2 & 0.17 & 0.15 &  59 \\
 23.0$-$23.5 & 129 & 1.05 & 0.24$\pm0.02$ & 1.0 & 0.24 & 0.19 & 150 \\
 23.5$-$24.0 &  36 & 0.77 & 0.20$\pm0.08$ & 0.8 & 0.29 & 0.22 & 212 \\
 24.0$-$24.5 &   4 & \nodata & \nodata & \nodata & \nodata & & 201 \\
 24.5$-$25.0 &   0 & \nodata & \nodata & \nodata & \nodata & &  52 \\
 25.0$-$25.5 &   0 & \nodata & \nodata & \nodata & \nodata & &   2 \\
\hline
\end{tabular}
\end{center}

\subsection{Estimate of the Metallicity Spread}

We now compare our IR CMD with the observed cluster giant branches and
theoretical isochrones from \citet{bertelli94} and BC00 to estimate the
upper and lower bound on the metallicity distribution of the stars in
the halo of NGC~5128.

We compare our data with the red giant branches of the clusters M92,
47~Tuc, M67, and NGC~6553 in Figure~\ref{cm_gal_charlot}.  The
references for the cluster data, their distance moduli, and their
metallicities are listed in Table~3.  The star cluster magnitudes were
transformed to [F160W] and [F110W] magnitudes according to the formulae
given in Appendix~A.  The NGC~6553 giant branch of \citet{davidge94} was
corrected for reddening ($E(J-H)=0.231$) and extinction ($A_H=0.38$)
\citep{rieke85,guarnieri98}.  The metallicity bounds on the data are
chosen to bracket $\pm1\sigma$ of the real color spread.  The most metal
poor stars in this halo field have metallicities matching those of the
globular cluster M92 with [Fe/H]=$-2.03$.  This is in good agreement
with the lower metallicity bound derived by SMW96 and HHP99.  At the
other end of the metallicity range NGC~6553, with [Fe/H]=$-0.29$,
reproduces well the upper bound of our IR data as for the optical data
of HHP99.  The cluster M67 is consistent with being more metal rich than
47~Tuc but its low magnitude giant branch restricts us from using it as
a comparison to our data.  From this comparison we conclude that the
metallicity spread based on the color spread in the IR halo data is
$-2.0\lesssim$[Fe/H]$\lesssim-0.3$ (these limits cover slightly more
than $\pm1\sigma$ of the real color spread; see
Figure~\ref{cm_gal_charlot}).  This comparison confirms the presence of
a metallicity spread as estimated in Section~6.1.

Assuming an old population, the IR data support the findings of SMW96
and HHP99 in the optical suggesting that the halo of NGC~5128 is
composed of stars ranging from metal-poor to near-solar metallicities.
To complicate things, the presence of an intermediate-age population, as
will be discussed in Section~6.4, can be another contributor to the
large spread in metallicities at magnitudes below the TRGB where the
populations become intertwined.  In fact, it is worth noting that
previous work, including the analysis presented here, has not solved for
the age of the population but assumed it to be old in order to solve for
the metallicity spread. If we assume that the spread in metallicity is
indeed associated with an old population, then this spread suggests that
metal enrichment occurred during the primordial collapse of the galaxy,
or alternative, that a low-metallicity component was accreted from an
external source.  Based on the hierarchical picture of galaxy formation,
HHP99 proposed that the old metal-poor stars in the halo formed during a
first burst of star formation occurring in the in-falling clumps of gas.
A first wave of supernovae explosions then ejected and enriched the gas
that eventually fell back into the potential well ($\sim 1-2$ Gyr later)
to form the more metal-rich stars in the halo.

The comparison with star cluster data is the best evidence of a spread
in metallicity in the old population of NGC~5128.  Because of the model
uncertainties, comparison with theoretical isochrones can only be used
as an example of how to tie the spread in colors to a spread in absolute
metallicity.  We compare our data with the 12 Gyr old stellar population
isochrones of \citet{bertelli94} in Figure~\ref{cm_bert_charlot}.  For a
meaningful comparison, the magnitudes were properly transformed using
the mathematical expressions derived in Appendix~A. The isochrones shown
in Figure~\ref{cm_bert_charlot} are for the metallicities [Fe/H]=$-1.7$,
$-0.7$, $-0.4$, $+0.0$, and $+0.4$.  The upper parts of these IR
isochrones with [Fe/H]$\geq-1.7$ (e.g., [F160W]~$<$~22.19 or
$M_{H_{BB}}<-5.59$ for [Fe/H]=$-0.7$; $BB$ refers to the Bessell and
Brett (1988) system; see Appendix~A and Figure~\ref{transform}) are not
yet satisfactory according to \citet{bertelli99}, due to the problems in
atmosphere models and scales of effective temperature for M giants.  The
comparison shows that our data are clearly more metal rich than the
[Fe/H]=$-1.7$ isochrone and at least as metal rich as [Fe/H]=$-0.7$.
Unfortunately, we cannot put a stronger upper limit on the spread in
metallicity because of the problems described above.

We also make use of the most up-to-date stellar evolution models of BC00
to interpret the color spread in our CMD.  The new isochrones of BC00
are generated from the new library of metallicity-dependent spectra
calibrated by \citet{lejeune97,lejeune98} and an improved
color-effective temperature relation for cool stars.  In addition, the
BC00 isochrones include the carbon AGB stars (see BC00 and Liu, Charlot
and Graham 2000 for details).  The prescriptions for those are
semi-empirical, and essentially based on models and observations of
stars in the Small Magellanic Cloud (SMC), Large Magellanic Cloud (LMC),
and Milky Way.  The BC00 isochrones indicate more clearly a mean
metallicity of [Fe/H]$\sim-0.4$ to $-0.7$, depending on magnitude (see
Figure~\ref{color_gauss}), and a range of
$-1.7\lesssim$[Fe/H]$\lesssim+0.0$, where the lower limit is clearly too
low for the $\pm1\sigma$ color spread.  The theoretical isochrones
indicate a slightly more metal-rich spread in metallicities than the
estimate from cluster giant branches.  This is also observed in the
optical data of HHP99.  The red upper limit indicates that the most
metal rich stars in the halo field we observed with NICMOS are roughly
solar in metallicity.

We use the 12~Gyr isochrones of BC00 to estimate the metallicity for
individual stars in our NICMOS field.  The metallicity histogram is
therefore only derived for stars with [Fe/H]=$-1.7$ to $+0.4$.  The
histogram shown in Figure~\ref{metal_table} for each 0.5 magnitude bin
in the range [F160W]=$21.0-24.0$ was generated for stars which fell on
or in between isochrones and for which we could estimate the metallicity
by using a simple linear interpolation method.  This is a very crude and
model dependent estimate of the metallicity, as it does not include
stars above the 12~Gyr isochrone TRGBs (e.g., bright intermediate-age
AGB stars) and ignores the stars redder (e.g., post-AGB stars in their
superwind phase) and bluer than the predicted colors (see
Figure~\ref{metal_table}).  The metallicity distribution for a total of
353 stars in our CMD, shown in Figure~\ref{metal_table_all}, peaks at
[Fe/H]=$-0.76$ with a dispersion of $\sigma=0.44$. This peak and
dispersion agree well with the overall metallicity distribution of the
globular clusters and of the halo stars of HHP99 (see
Figure~\ref{metal_table_all}; Harris et al. 1992; HHP99).  The [Fe/H]
distribution of stars in our NICMOS field does not seem to be resolved
into multiple peaks, as is the case for the halo stars of HHP99.  We
find no obvious sign of a ``sub-peak'' which would match the largest
sub-peak of the halo metallicity distribution of HHP99 at [Fe/H]=$-0.32$
with a dispersion of $\sigma=0.22$, corresponding nicely with the second
largest sub-peak in the globular cluster metallicity distribution.  It
is not clear at this point if the difference in the shape and peak of
the metallicity distribution between the IR and the optical data of
HHP99 is real or due to the large uncertainty associated with the
color-metallicity transformation.  The general agreement of the
metallicity spread is promising; the combination of the NICMOS
observations and the F555W and F814W observations of SMW96 of the same
halo field should help resolve the issue \citep{marleau00}.
  
\begin{center}
\begin{tabular}{c}
TABLE 3 \\ \small{ADOPTED PARAMETERS FOR M92, 47~TUC, M67, AND NGC~6553}
\end{tabular}
\end{center}
\vspace{-0.6cm}
\begin{center}
\begin{tabular}{ccccccc} 
\hline \hline
  Star Cluster & Cluster Type & CMD Data & [Fe/H] & $(m-M)_0$ & Magnitude System & References \\
\hline
  M92 & Globular & 4 & $-2.03$ & 14.6 & CIT & 1 \\
  47 Tuc & Globular & 7 & $-0.65$ & 13.4 & CIT & 2 \\
  M67 & Old Open & 3 & $-0.09$ & 9.38 & CIT & 3,4 \\
  NGC 6553 & Globular & 5 & $-0.29$ & 13.6 & CIT & 5,6 \\
\hline
\end{tabular}
\end{center}
\vspace{-0.6cm}
\begin{center}
\begin{tabular}{c}
\footnotesize{NOTES TO TABLE 3} \\
\end{tabular}
\end{center}
\vspace{-1.0cm}
\begin{center}
\begin{tabular}{l}
(1) \citet{stetson88}; (2) \citet{hesser87}; (3) \citet{houdashelt92}; \\
(4) \citet{cohen78}; (5) \citet{davidge94}; (6) \citet{guarnieri97}; \\
(7) \citet{frogel81} \\
\end{tabular}
\end{center}

\subsection{Bright Stars}

As can be seen in Figure~\ref{cm_gal_charlot} and Table~2, we detect a
population of bright stars above the TRGB of globular clusters and old
stellar populations.  These bright stars in our CMD can be due to (1)
blended images, (2) Galactic contamination, or (3) a young or
intermediate-age halo population.

We searched for blended doubles by first examining the {\sl SHARPNESS}
parameter calculated by DAOPHOT.  The {\sl SHARPNESS} is defined so that
is zero for stars, larger than zero for galaxies or unrecognized blended
doubles, and less than zero for cosmic rays or single pixel defects.  We
plotted the {\sl SHARPNESS} and $\chi^2$ as a function of magnitude for
our data and found the {\sl SHARPNESS} parameter to be close to zero for
those bright stars above the TRGB, with good values of $\chi^2$ (as
expected from our initial $\chi^2$ cut).  Therefore, we found no reason
to reject them.

The five brightest, blue stars in the upper left region of
Figure~\ref{cm_lmc_charlot} with [F160W]~$<$~20.0 and
0.40~$<$~[F110W]$-$[F160W]~$<$~0.85, have colors and magnitudes that are
consistent with foreground Galactic stars.  The theoretical isochrone of
\citet{bertelli94} for a 5 Gyr disk population with solar metallicity is
shown in Figure~\ref{cm_lmc_charlot} assuming a distance modulus for the
Galactic stars of 15.0.  Main sequence K$-$M dwarfs have $0.4 \leq J-H
\leq 0.7$ \citep{tokunaga95,davidge98} and therefore the five bright
stars have colors and magnitudes consistent with K$-$M dwarfs in our
Galaxy.  To assess the number of foreground Galactic stars expected to
contaminate our data, we used the Galaxy model of \citet{cohen93} for IR
star counts for the direction $(l,b)=(309.5^\circ, 19.4^\circ)$,
adopting a solar displacement of 15 pc, and a halo scale factor of 0.5
\citep{cohen95}.  Table~4 lists the predicted number of stars for 1
magnitude wide bins for the [F160W] filter.  The expected number of
foreground stars contaminating each magnitude bin in the F160W image
based on Cohen's model, scaled by a factor of 3.9, is shown in
Figure~\ref{lf_all_match}.  No correction was made between H and
[F160W].

\begin{center}
\begin{tabular}{c}
TABLE 4 \\
\small{PREDICTED NUMBER OF GALACTIC STARS} 
\end{tabular}
\end{center}
\vspace{-0.75cm}
\begin{center}
\begin{tabular}{cccc} 
\hline \hline
[F160W] & N$_p$ & N$_p \times$ 3.9 & N$_{obs}$ \\
\hline 
  17.0$-$18.0 &  0.30 &  1.17 &   1 \\
  18.0$-$19.0 &  0.42 &  1.64 &   1 \\
  19.0$-$20.0 &  0.55 &  2.14 &   3 \\
  20.0$-$21.0 &  0.68 &  2.65 &  30 \\
  21.0$-$22.0 &  0.78 &  3.04 & 134 \\
  22.0$-$23.0 &  0.81 &  3.16 & 328 \\
  23.0$-$24.0 &  0.74 &  2.89 & 165 \\
  24.0$-$25.0 &  0.60 &  2.34 &   4 \\
  25.0$-$26.0 &  0.43 &  1.68 &   0 \\
  26.0$-$27.0 &  0.28 &  1.09 &   0 \\
\hline
\end{tabular}
\end{center}

Since the bright stars ([F160W]~$<$~20.0) in the NIC2 image have colors
consistent with Galactic dwarfs, it seems likely that Cohen model
under-predicts the contamination since it predicts only 1.27 stars in
the range [F160W]=17.0--20.0.  The Cohen model has not been validated at
these faint magnitudes and therefore we feel justified in applying an
arbitrarily scaling to the Cohen predictions.  A maximum-likelihood
analysis shows that scaling up the predicted counts ($N_p$) by a factor
of 3.9 produces the best fit to the observed counts ($N_{obs}$) with a
95\% confidence interval of $2.1-8.3$.  With this scale factor we
predict the contamination in fainter magnitude bins as given in Table~4.
Clearly, this implies that Galactic contamination is insignificant for
all bins fainter than [F160W]=20.0.

\subsection{The Intermediate-Age Population}

The presence of an intermediate-age population of $\sim5$~Gyr in the
halo of NGC~5128 was proposed by SMW96 based on their detection of
$\sim200$ stars brighter than the TRGB.  We look for the presence of
intermediate-age stars as postulated by SMW96 by comparing the IR data
with the 2 Gyr old isochrone from BC00 with metallicity [Fe/H]=$-1.7$,
$-0.7$, $-0.4$, $+0.0$, and $+0.4$, which include the carbon AGB stars.
Figure~\ref{cm_lmc_charlot} shows that the presence of an
intermediate-age population would be revealed by asymptotic giant branch
(AGB) stars $\sim1$ mag above the TRGB.  This effect is also
demonstrated by moving the LMC and SMC AGB stars with ages between $1-3$
Gyr and the younger stars with ages between $40-120$ Myr to the distance
of NGC~5128 \citep{frogel90}.  The AGB is composed of M-type
(oxygen-rich) stars and, for young- and intermediate-age populations, of
C-type (carbon-rich) stars which occupy the bright end of the AGB.

As Figure~\ref{cm_lmc_charlot} shows, the LMC and SMC AGB stars
belonging to an intermediate-age population occupy a part of the CMD
that is also occupied by the brightest IR stars in the halo of NGC~5128.
We find that $\sim10$\% of the stars resolved in our NICMOS images and
appearing in our CMD are brighter than the TRGB of a 12~Gyr old
population, assuming a metallicity of [Fe/H]=$+0.0$ (this brightest
isochrone TRGB terminates at [F160W]=21.2 or $M_{H_{BB}}=-6.6$; BC00).
The IR data suggest the presence of an intermediate-age population, also
seen in the WFPC2 observations of SMW96.  A preliminary match between
the IR and the SMW96 optical data shows that our intermediate-age
population consists of the same stars that make up the intermediate-age
population of SMW96 \citep{marleau00}.  For their WFPC2 field, located
18\myarcminpoint32 south of the galaxy center, HHP99 claim that the halo
is composed almost entirely of old stars (at most 300 stars,
representing $\sim3$\% of their sample, belong to an intermediate-age
population).  As the fraction of intermediate-age stars detected in the
WFPC2 SMW96 field, in agreement with our NICMOS field, is larger by at
least a factor of $\sim3$ at half the radial distance from the galaxy
center, it is suggestive of a radial gradient in the stellar population
in the halo (the younger population being more centrally concentrated).
With the limited area of our NICMOS field and hence the small number
statistics, it is not possible to accurately estimate the metallicity
mean and spread associated with the intermediate-age population only.
Assuming an old halo population, the double metallicity peak in the
globular cluster systems of NGC~5128 \citep{harris92} and the halo stars
in the HPP99 field at [Fe/H]$\simeq-1.1$ (barely noticeable for the halo
stars) and $-0.3$ is well within our estimated range in metallicities.

If one excludes the brightest star at [F160W]=17.15 which is most
certainly due to Galactic contamination, the colors and magnitudes of
the other four stars with [F160W]~$<$~20.0 are close to being consistent
with a very young ($40-120$ Myr) population (see
Figure~\ref{cm_lmc_charlot}).  But since their colors are closer to the
late type K$-$M dwarfs which dominate Galactic contamination, we conclude
that there is no strong evidence for a very young population of stars in
our halo field. 

\section{Conclusion} 

We have presented the first IR CMD for the halo of a giant elliptical
galaxy.  Assuming a distance to NGC~5128 of 3.5 Mpc, we have detected a
discontinuity in the luminosity function at [F160W]$\approx$20.0 and
have measured IR magnitudes and colors for stars in the halo of NGC~5128
to [F160W]=23.8 (50\% completeness limit).  We are confident that we are
not confused by crowding to [F160W]$\simeq23.5$ based on careful
analysis of artificial-stars tests.  The weighted average of the mean
color of our giant branch above our 50\% completeness limit is
[F110W]$-$[F160W]=1.22$\pm0.08$ ($(J-H)_{CIT}=0.78$) with a dispersion
of 0.19 mag.  From our artificial-star experiments we have determined
that there is a real spread in color in our CMD.  By comparing our data
with star cluster giant branches and theoretical isochrones, we were
able to constrain the metallicity spread associated with this real color
spread.  Assuming an old population, we find that, in the halo field of
NGC~5128 we surveyed, stars have metallicities ranging from roughly 1\%
of solar at the blue end of the color spread to roughly solar at the red
end, with a mean of [Fe/H]=$-0.76$ and a dispersion of 0.44 dex.

We assert that the five brightest stars above the SMW96 determination of
the TRGB are most probably due to Galactic contamination.  We found that
the majority of stars above the TRGB of an old population belong to an
intermediate-age population ($\sim2$~Gyr).  The presence of an
intermediate-age population in the halo of NGC~5128 is consistent with
the findings of SMW96.  We conclude from our analysis that the IR data
are consistent with the halo of NGC~5128 being composed of at least two
age populations, a population with ages $\sim2$~Gyr and an old
population.  Assuming an old population, we find that the stars have a
wide range of metallicities.  Future work will combine the WFPC2 CHIP-3
observations of SMW96 and our NICMOS data to examine the multi-color
(F555W, F814W, F110W, and F160W) properties of stars in the halo of
NGC~5128 and try to reconstruct the galaxy's formation history.

\acknowledgments F.R.M. would like to thank Nial Tanvir for providing
some of the software for the completeness tests and Rachel Johnson for
her instructions on aperture photometry corrections.  Many thanks to Jay
Anderson for the use of his code with which he initially derived for us
the WFPC2(SMW96) and NICMOS star matching.  We are grateful to Joan
Najita and Patricia Royle for private communications while dealing with
cosmic rays persistence effects and data reduction related issues.
Finally, we would like to mention that Martin Cohen kindly provided us
with the star count calculations from his model.  This research was
supported by the HST NASA grant STScI GO-07852.02-96A.
F.R.M. acknowledges an IoA observational astronomy rolling grant from
PPARC, ref. no. PPA/G/O/1997/00793.

\appendix
\section{Magnitude System Transformations}

The filters used for the NICMOS observations differ substantially from
ground-based IR systems.  The central wavelength of F110W is close to
that of $J$-band (1.3 $\mu$m), but this filter is almost twice as wide
as $J$ and transmits from $0.8-1.4 \mu$m.  The F160W is analogous to
ground-based $H$, but extends an additional 0.1 $\mu$m blueward.  To
facilitate comparison of our NICMOS data with previous ground-based IR
observations, and with theoretical isochrones, we have derived color
transformations between F110W and F160W and their closest ground-based
counterparts, $J$ and $H$.

We need to compare our data with stellar cluster photometry (all of
which is on the CIT system) and with the isochrones of 
\citet{bertelli94} which are presented on the homogenized system proposed by 
\citet{bessell88}. They present an empirical transformation
from CIT to their system equal to:

\[ (J-H)_{BB} = 0.002 + 1.098~(J-H)_{CIT}. \]

\noindent 
Therefore, we concentrate on calculating the transformation between the
BB and NICMOS system.  This transformation must be deduced theoretically
because only a small number of calibration stars (five) were observed
both by NICMOS (see NICMOS calibration web page at STScI) and from the
ground \citep{persson79}. Eventually, when more NICMOS data become
public it will be possible to replace this theoretical transformation
with an empirical one.

We have calculated synthetic magnitudes for stars in the \citet{pickles98}
spectral library in the NICMOS and BB systems. Since we have adopted a
Vega-based system all colors are relative to that of an A0V star, i.e.,
if $\eta_\lambda$ is the system efficiency, taken from \citet{bessell88}
and from the HST Data Handbook version 3.1 March 1998
respectively, and $f_\lambda$ is the stellar flux, the corresponding
synthetic magnitude $m_\lambda$ is:

\[ m_\lambda = -2.5 \; log \left(
{\int_0^\infty f_\lambda \eta_\lambda d\lambda \over \int_0^\infty
f(A0V)_\lambda \eta_\lambda d\lambda} \right). \]

\noindent
Figure~\ref{transform} shows an example of synthetic Vega-based colors
for the Pickles library stars on the BB and NICMOS systems.  The
spectral types of the giants are O8III to M10III and O5V to M6V for
dwarfs; both normal, and a few metal-poor and metal-rich stars have been
included.  This figure shows that there is a significant color
difference between [F110W]$-$[F160W] and $(J-H)_{BB}$.  A simple linear
fit,

\[ [F110W]-[F160W] = -0.058 + 1.484 (J-H)_{BB}, \]

\noindent 
and

\[ [F160W] = 0.019 + 0.095 (J-H)_{BB} + H_{BB}, \]

\noindent
describes the transformation accurately.  There is some scatter about
the straight line for for the latest M giants. Given the mean colors of
the stars in NGC~5128 we expect any systematic error to be less than 0.1
mag in locating the transformed isochrones.  An error of 0.1 mag in
[F110W]$-$[F160W] corresponds to a 0.2 dex error in metallicity.

The colors for five stars observed as part of the NICMOS photometric
calibration campaign with magnitudes measured in both the NICMOS and Las
Campanas Observatory (LCO) \citep{persson79} systems are also plotted
in Figure~\ref{transform}.  The LCO magnitudes of these stars were first
transformed to the BB system.  Comparison of the data with the synthetic
photometry in Figure~\ref{transform} confirms that the theoretical
colors are satisfactory and that the transformation can be made reliably
for normal stars. Some of Persson's stars are highly reddened and fall
off the linear trend.  However, nearly all of our stars have
[F110W]$-$[F160W]~$<$~1.5, and so the quadratic term that emerges for very
red stars is not relevant.

\clearpage

\figcaption[ngc5128_dss_15.ps]{The HST field of view (FOV) of the NICMOS
cameras ({\it solid squares}) overlaid on the Digital Sky Survey image
15\myarcminpoint0 $\times$ 15\myarcminpoint0 wide. From left to right,
the {\it solid squares} represent NIC3, NIC1 and NIC2.  The WFPC2 FOV of
SMW96 is shown as the {\it dotted squares}.  The geometric center of the
NIC1+NIC2 combined FOV was chosen to coincide with the geometric center
of the WFPC2 CHIP-3 FOV of SMW96.  The NIC1+NIC2 FOV was chosen so that
it falls inside the WFPC2 CHIP-3 FOV for any position angle.
\label{ngc5128_dss_15}}

\figcaption[ngc5128_nic2.ps]{NIC2 F110W ({\it top}) and F160W ({\it
bottom}) mosaics.  The {\it left} images show the reduced NIC2 mosaics
in each filter.  Each NIC2 mosaic covers a field of view
20\myarcsecpoint4 $\times$ 20\myarcsecpoint4 wide.  The labels are in
arcseconds and the grey scale represents counts per second.  The {\it
right} images show a 10\myarcsecpoint2 $\times$ 10\myarcsecpoint2 wide
close up of the central region of the mosaics.  The Airy rings are
clearly visible around the unblended stars.
\label{ngc5128_nic2}}

\figcaption[matches.ps]{ The histogram of number of matches between
stars in the F110W and F160W images. Poisson error bars are plotted.
The {\it solid line} is the adopted model describing the matching
process, which assumes a Gaussian distribution of centroid errors and a
uniform background of random matches ({\it dashed line}).  Inspection of
this histogram suggests that a natural cut off for the matching radius
is 1 pixel. The model then indicate that 3\% of our sample is matched
with the wrong star.
\label{matches}}

\figcaption[completeness.ps]{The completeness functions for the F160W
and F110W NIC2 observations. Completeness tests were performed by adding
artificial stars to each individual image.  Stars were simulated with
magnitudes between 20.0 and 26.5, at a 0.1 mag interval.  Each
simulation consisted of adding 132 stars to the real image,
corresponding to 10\% of the stars recovered from the F160W image.  The
50\% completeness level occurs at [F110W]=24.5 and [F160W]=23.8,
respectively.  \label{completeness}}

\figcaption[lf_all_match.ps]{{\it At the top:} The luminosity functions
for stars detected in the F110W ({\it dashed lines}) and F160W ({\it
solid lines}) image respectively, taking into account the edge and
$\chi^2$ cuts. {\it At the bottom:} The luminosity functions for each
respective color image obtained after applying the matching criterion
($r_m=1$ pixel).  The {\it dotted} histogram labeled {\it Galactic}
represents the expected number of foreground stars contaminating each
magnitude bin in the F160W image based on the Galaxy model of
\citet{cohen93} scaled by a factor of 3.9 (see Table~4).
\label{lf_all_match}}

\figcaption[transform.ps]{Magnitude systems transformations.  The plot
shows our adopted linear transformation ({\it solid line}) between the
homogenized system of \citet{bessell88} and NICMOS. The {\it filled
circles} and {\it open lozenges} are synthetic magnitudes calculated on
a Vega-based system for the dwarf and giant stars respectively in the
\citet{pickles98} spectral library.  The {\it crosses} are the data for
stars from the NICMOS photometric calibration campaign (STScI NICMOS
calibration web page) that also have ground based measurements from
\citet{persson79}.  The observational errors are smaller than the cross
symbols used to plot the points. The \citet{persson79} data have been
transformed from the original Las Campanas Observatory system to the
Bessell and Brett system (BB).
\label{transform}}

\figcaption[cm_mean.ps]{The plot shows {\it on the left} the data points
and the mean (points) and standard deviation (error bars) for each
magnitude interval.  The dotted line {\it on the right} shows the total
uncertainties measured from our artificial-star experiments.  The lines
join the mean value at each magnitude.  The vertical error bars
represent the mean of the DAOPHOT photometric errors for both the real
data and the artificial-star tests.  The {\it dashed line} defines the
50\% completeness limit.  The reddening vector has an amplitude of
$E(J-H)=0.04$ and $A_H=0.06$ (smaller than the data points).
\label{cm_mean}}

\figcaption[color_gauss.ps]{For each 0.5 magnitude bin in the range
[F160W]=$20.5-24.0$, histogram showing the distribution of
[F110W]$-$[F160W] color for our NICMOS data.  Overplotted on each
histogram is a Gaussian fit using the mean and standard deviation
calculated for each magnitude bin.
\label{color_gauss}}

\figcaption[cm_gal_charlot_BW.ps]{{\it On the left:} Comparison of our
data ({\it filled circles}) with the 12 Gyr old stellar population
isochrones of BC00.  From left to right, BC00's 12 Gyr isochrone for the
metallicity [Fe/H]=$-1.7$, $-0.7$, $-0.4$, $+0.0$, and $+0.4$.  Each
individual {\it solid line} represents the RGB and each {\it dotted line}
shows the AGB phase.  {\it On the right:} CMD comparing our data to the
red giant branches of globular clusters.  From left to right, the {\it
solid lines} are the loci of the giant branches of the four clusters
M92, 47~Tuc, M67 and NGC~6553 with metallicities [Fe/H]=$-2.0$, $-0.7$,
$-0.1$, and $-0.3$, respectively.
\label{cm_gal_charlot}}

\figcaption[cm_bert_charlot_BW.ps]{{\it On the left:} Comparison of our
data ({\it filled circles}) with the 12 Gyr old stellar population
isochrones of \citet{bertelli94} (RGB only).  From left to right, the 12
Gyr isochrone for the metallicity [Fe/H]=$-1.7$, $-0.7$, $-0.4$, $+0.0$,
and $+0.4$.  The upper parts of the [Fe/H]$\geq-1.7$ $(J-H)_{BB}$
isochrones (e.g., [F160W]~$<$~22.19 or $M_{H_{BB}}<-5.59$ for
[Fe/H]=$-0.7$) are not yet satisfactory according to \citet{bertelli99},
due to the problems in atmosphere models and scales of effective
temperature for M giants.  {\it On the right:} Comparison with the 12
Gyr old stellar population isochrones of BC00. Each individual {\it
solid line} represents the RGB whereas each {\it dotted line} shows the
AGB. \label{cm_bert_charlot}}

\figcaption[metal_table.ps]{For each 0.5 magnitude bin in the range
[F160W]=$21.0-24.0$, histogram showing the distribution of [Fe/H] for
the stars in our CMD with [F160W] magnitudes fainter than the TRGBs of
the 12 Gyr isochrones of BC00 (metallicity dependent; the brightest TRGB
is at [F160W]=21.2) and with [F110W]$-$[F160W] colors within the model
predictions.  The metallicities of individual stars can only be
estimated for stars with [Fe/H]=$-1.7$ to $+0.4$ by using the 12 Gyr
isochrones of BC00 and a simple linear interpolation method.
\label{metal_table}}

\figcaption[metal_table_all.ps]{{\it Top panel:} Histogram showing the
distribution of [Fe/H] for the 353 stars in our CMD with [F160W]
magnitudes fainter than the TRGBs of the 12 Gyr isochrones of BC00
(metallicity dependent; the brightest TRGB is at [F160W]=21.2) and with
[F110W]$-$[F160W] colors within the model predictions.  The peak of the
Gaussian fit, over-plotted on the histogram, is at [Fe/H]=$-0.76$ with a
dispersion of $\sigma=0.44$.  {\it Middle panel:} The [Fe/H]
distribution for the brightest halo field stars of HHP99 with magnitudes
$I=25.0-26.0$.  The curve represents the Gaussian fit to our NICMOS data
that has been re-normalized to the halo stars counts.  {\it Bottom
panel:} The [Fe/H] distribution for 47 globular clusters in the halo of
NGC~5128 ($R<$~4\arcmin; Harris et al. 1992). The Gaussian curve fitted
to our NICMOS data is over-plotted on the histogram, re-normalized to the
globular cluster counts.  The dispersion of the NICMOS halo stars [Fe/H]
distribution agrees well with the overall metallicity dispersion of the
globular clusters and the halo stars of HHP99.  The overall distribution
does not seem to be resolved into multiple peaks, as is the case for the
halo stars of HHP99 and the globular clusters.
\label{metal_table_all}}

\figcaption[cm_lmc_charlot_BW.ps]{{\it On the left:} Comparison of our
data ({\it filled circles}) with the 2 Gyr old stellar population
isochrones of BC00.  Each individual {\it solid line} represents the RGB
whereas each {\it dotted line} shows the AGB.  From left to right, the 2
Gyr isochrone for the metallicity [Fe/H]=$-1.7$, $-0.7$, $-0.4$, $+0.0$,
and $+0.4$.  The last point on the isochrones with metallicity
[Fe/H]=$-0.7$, $-0.4$, and $+0.0$ is for the carbon AGB stars.  The {\it
long dashed} line represents a 5 Gyr old isochrone with solar
metallicity, assuming a distance modulus of 15.0 for the Galactic stars
contamination.  The colors and magnitudes of the 5 brightest stars above
the TRGB are consistent with being Galactic K$-$M dwarfs along our line
of sight.  {\it On the right:} Comparison of LMC+SMC AGB stars with our
data.  This shows that the presence of a young- to intermediate-age
population would be revealed by stars $1-2$ mag above our observed TRGB.
The {\it small diagonal crosses} and {\it plus symbols} are LMC+SMC AGB
M-type and C-type stars, respectively, with ages between $1-3$ Gyr.  The
{\it large diagonal crosses} and {\it plus symbols} are LMC+SMC AGB
M-type and C-type stars, respectively, with ages between $40-120$ Myr.
\label{cm_lmc_charlot}}


\begin{thebibliography}{}
\bibitem[Alonso and Minniti(1997)] {alonso97} Alonso, M.V., and Minniti, D. 1997, \apjs, 109, 397
\bibitem[Bertelli(1999)] {bertelli99} Bertelli, G. 1999, private communication
\bibitem[Bertelli et al.(1994)] {bertelli94} Bertelli, G., Bressan, A., Chiosi, C., Fagotto, F., and Nasi, E. 1994, \aap, 106, 275
\bibitem[Bessell and Brett(1988)] {bessell88} Bessell, M.S., and Brett, J.M. 1988, \pasp, 100, 1134
\bibitem[Bruzual and Charlot(2000)] {bruzual00} Bruzual, A.G., and Charlot, S. 2000, 
in preparation (BC00)
\bibitem[Bruzual and Kron(1980)] {bruzual80} Bruzual, A.G., and Kron, R.G. 1980, \apj, 241, 25
\bibitem[Calzetti, Dickinson and Rieke(1999)] {calzetti99} Calzetti, D., Dickinson, M., and Rieke, M. 1999, private communication
\bibitem[Cohen(1995)] {cohen95} Cohen, M. 1995, \apj, 444, 874
\bibitem[Cohen(1993)] {cohen93} Cohen, M. 1993, \aj, 105, 1860 
\bibitem[Cohen, Frogel and Persson(1978)] {cohen78} Cohen, J.G., Frogel,
J.A., and Persson, S.E. 1978, \apj, 222, 165
\bibitem[Cole et al.(1994)] {cole94} Cole, S., Arag\'on-Salamanca, A., Frenk, C.S., Navarro, J.F., and Zepf, S.E. 1994, \mnras, 271, 781
\bibitem[Colina and Rieke(1997)] {colina97} Colina, L., and Rieke,
M.J. 1997, in {\it HST Calibration Workshop}, eds. S. Casertano et al.
\bibitem[Davidge(1998)] {davidge98} Davidge, T.J. 1998, \apj, 497, 650
\bibitem[Davidge and Simons(1994)] {davidge94} Davidge, T.J., and Simons, D.A. 1994, \aj, 107, 240
\bibitem[Dufour et al.(1979)] {dufour79} Dufour, R.J., van den Bergh, S., Harvel, C.A., et al. 1979, \apj, 84, 284
\bibitem[Frogel et al.(1990)] {frogel90} Frogel, J.A., Terndrup, D.M., Blanco, V.M., and 
Whitford, A.E. 1990, \apj, 353, 494
\bibitem[Frogel(1984)] {frogel84} Frogel, J.A. 1984, \apj, 278, 119
\bibitem[Frogel, Persson and Cohen(1981)] {frogel81} Frogel, J.A., Persson, S.E., and Cohen, J.G. 1981, \apj, 246, 842
\bibitem[Graham(1998)] {graham98} Graham, J. A. 1998, \apj, 502,  245
\bibitem[Guarnieri et al.(1998)] {guarnieri98} Guarnieri, M.D.,
Ortolani, S., Montegriffo, P., Renzini, A., Barbuy, B., Bica, E., and
Moneti, A. 1998, \aap, 331, 70
\bibitem[Guarnieri, Renzini and Ortolani(1997)] {guarnieri97} Guarnieri, M.D., Renzini, A., and Ortolani, S. 1997, \apj, 477, L21
\bibitem[Guiderdoni and Rocca-Volmerange(1990)] {guiderdoni90} Guiderdoni, B., and 
Rocca-Volmerange, B. 1990, \aap, 227, 362
\bibitem[Harris, Harris and Poole(1999)] {harris99} Harris, G.L.H, Harris, W.E., 
and Poole, G.B. 1999, \aj, 117, 855 (HHP99)
\bibitem[Harris et al.(1992)] {harris92} Harris, G.L.H., Geisler, D., Harris, H.C., and 
Hesser, J.E. 1992, \aj, 104, 613
\bibitem[Harris (1986)] {harris86} Harris, W.E. 1986, in The
Harlow Shapley Symposium on Globular Cluster Systems in Galaxies, IAU
Symposium No. 126, edited by J.E. Grindlay and A.G.D. Philip (Reidel,
Dordrecht), p. 237
\bibitem[Hesser et al.(1987)] {hesser87} Hesser, J.E., Harris, W.E.,
VandenBerg, D.A., Allwright, J.W.B., Shott, P., and Stetson, P.B. 1987, \pasp, 99, 739
\bibitem[Hibbard et al.(1994)]{hibbard94} Hibbard, J.E., Guhathakurta, P., van Gorkom, J.H.,
and Schweizer, F. 1994, \aj, 107, 67
\bibitem[Houdashelt, Frogel and Cohen(1992)] {houdashelt92} Houdashelt, M.L., Frogel, J.A., and Cohen, J.G. 1992, \aj, 103, 163
\bibitem[Hui et al.(1993)] {hui93} Hui, X., Ford, H.C., Ciardullo, R., and 
Jacoby, G.H. 1993a, \apj, 414, 463
\bibitem[Israel(1998)] {israel98} Israel, F.P. 1998, \aapr, 8, 237
\bibitem[Jablonka et al.(1996)] {jablonka96} Jablonka, P., Bica, E., Pelat, D., and 
Alloin, D. 1996, \aap, 307, 385
\bibitem[Kauffmann, White and Guiderdoni(1993)] {kauffmann93} Kauffmann, G., White, S.D.M., and Guiderdoni, B. 1993, \mnras, 264, 201
\bibitem[King and Ellis(1985)] {king85} King, C.R., and Ellis, R.S. 1985, \apj, 288, 456
\bibitem[Koo(1981)] {koo81} Koo, D.C. 1981, PhD thesis, University of California at Berkeley
\bibitem[Krist(1993)] {krist93} Krist, J. 1993, in {\it Astronomical Data Analysis
Software and Systems II}, 52, eds. R.J. Hanisch, R.J.V.
Brissenden and J. Barnes (ASP Conference Series), 536 
\bibitem[Lacey et al.(1993)] {lacey93} Lacey, C., Guiderdoni, B., Rocca-Volmegange, B., and 
Silk, J. 1993, \apj, 402, 15
\bibitem[Lejeune, Cuisinier and Buser(1998)] {lejeune98} Lejeune, T., Cuisinier, F., and 
Buser, R. 1998, \aaps, 130, 65
\bibitem[Lejeune, Cuisinier and Buser(1997)] {lejeune97} Lejeune, T., Cuisinier, F., and
Buser, R. 1997, \aaps, 125, 229
\bibitem[Liu, Charlot and Graham(2000)] {liu00} Liu, M.C., Charlot, S., and Graham,
J.R. 2000, \apj, in press
\bibitem[MacKenty et al.(1997)] {mackenty97} MacKenty, J. et al. 1997,
NICMOS Instrument Handbook (Baltimore: STScI)
\bibitem[Malin, Quinn and Graham(1983)] {malin83} Malin, D.F., Quinn, P.J., and 
Graham, J.A. 1983, \apj, 272, L5
\bibitem[Marleau et al.(2000)] {marleau00} Marleau, F.R., Graham, J.R.,
Liu, M.C., and Charlot, S. 2000, in preparation
\bibitem[Mathieu, Dejonghe and Hui(1996)] {mathieu96} Mathieu, A., Dejonghe, H.,
and Hui, X. 1996, \aap, 309, 30
\bibitem[Minniti et al.(1996)] {minniti96} Minniti, D., Alonso, M.V.,
Goudfrooij, P., Jablonka, P., and Meylan, G. 1996, \apj, 467, 221
\bibitem[Najita, Dickinson and Holfeltz(1998)] {najita98} Najita, J.,
Dickinson, M., and Holfeltz, S. 1998, Instrument Science Report NICMOS 98001
\bibitem[Persson, Frogel and Aaronson(1979)] {persson79} Persson, S.E., Frogel, J.A., and 
Aaronson, M. 1979, \apjs, 39, 61 
\bibitem[Pickles(1998)] {pickles98} Pickles, A.J. 1998, \pasp, 110, 863
\bibitem[Quillen, Graham and Frogel(1993)] {quillen93} Quillen, A.C., Graham, J.R., and 
Frogel, J.A. 1993, \apj, 412, 550
\bibitem[Renzini(1998)] {renzini98} Renzini, A. 1998, \aj, 115, 2459
\bibitem[Rieke(1999)] {rieke99} Rieke, M. April 1999, private communication
\bibitem[Rieke and Lebofsky(1985)] {rieke85} Rieke, G. H., and Lebofsky, R. M. 1985, \apj, 288, 618
\bibitem[Schiminovich et al.(1994)] {schiminovich94} Schiminovich, D., van Gorkom, J.H., 
van der Hulst J.M., and Kasow, S. 1994, \apj, 423, L101
\bibitem[Shanks et al.(1984)] {shanks84} Shanks, T., Stevenson, P.R.F., Fong, R., and 
McGillivray, H.T. 1984, \mnras, 206, 767
\bibitem[Silva and Bothun(1998)] {silva98} Silva, D.R., and Bothun, G.D. 1998, \aj, 116, 85
\bibitem[Somerville(1997)] {somerville97} Somerville, R. 1997, PhD thesis, University of California at Santa Cruz
\bibitem[Soria et al.(1996)] {soria96} Soria, R., Mould J.R., Watson A.M., 
et al. 1996, \apj, 465, 79 (SMW96)
\bibitem[Stetson(1992)] {stetson92} Stetson, P.B. 1992, in {\it Astronomical Data Analysis Software and Systems I}, eds. D.M. Worrall, C. Biemesderfer, and J. Barnes (San Francisco ASP), ASP Conference Series, 25, 297
\bibitem[Stetson and Harris(1988)] {stetson88} Stetson, P.B., and Harris, P.B. 1988, \aj, 96, 909
\bibitem[Stetson(1987)] {stetson87} Stetson, P.B. 1987, \pasp, 99, 191
\bibitem[Tinsley(1980)] {tinsley80} Tinsley, B.M., 1980, \apj, 241, 41
\bibitem[Tinsley and Gunn(1976)] {tinsley76} Tinsley, B.M., and Gunn, J.E. 1976, \apj, 203, 52
\bibitem[Thompson et al.(1998)] {thompson98} Thompson, R.I., Rieke, M., Schneider, G., Hines, D.C, and Corbin, M.R. 1998, \apj, 492, L95
\bibitem[Tokunaga(1995)] {tokunaga95} Tokunaga, A.T. 1995, Infrared
Astronomy Chapter for a revised Astrophysical Quantities, Cox, A.N., Editor.
\bibitem[Tonry et al.(1997)] {tonry97} Tonry, J.L., Blakeslee, J.P., Ajhar,
E.A., and Dressler, A. 1997, \apj, 475, 399
\bibitem[Tonry(1991)] {tonry91} Tonry, J.L. 1991, \apj, 373, L1
\bibitem[Tonry and Schechter(1990)] {tonry90} Tonry, J.L., and Schechter P.L. 1990, \aj, 100, 1794
\bibitem[Toomre and Toomre(1972)]{toomre72} Toomre, A. and Toomre, J. 1972, \apj, 178, 623
\bibitem[Tubbs(1980)] {tubbs80} Tubbs, A.D. 1980, \apj, 241, 969
\bibitem[van den Bergh(1976)] {vandenbergh76} van den Bergh, S. 1976, \apj, 208, 673 
\bibitem[Weil and Hernquist(1996)] {weil96} Weil, M.L., and Hernquist, L. 1996, \apj, 460, 101
\bibitem[White and Frenk(1991)] {white91} White, S.D.M., and Frenk, C.S. 1991, \apj, 379, 52
\bibitem[e.g., White and Rees(1978)] {white78} White, S.D.M., and Rees, M.J. 1978, \mnras, 183, 341
\bibitem[e.g., Worthey, Faber and Gonzalez(1992)] {worthey92} Worthey, G., Faber, S.M, and 
Gonzalez, J.J 1992, \apj, 398, 69
\bibitem[Yoshii and Takahara(1988)] {yoshii88} Yoshii, Y. and Takahara, F. 1988, \apj, 326, 1
\end{thebibliography}
\end{document}